\documentclass{mn2e}
\usepackage{amsfonts}
\usepackage{amsmath}
\usepackage{amssymb}
\usepackage{graphicx}

\newcommand{\p}{\ensuremath{\partial}}
\newcommand{\del}{\ensuremath{\delta}}
\newcommand{\Del}{\ensuremath{\Delta}}

\newcommand{\ep}{\ensuremath{\epsilon}}
 
\newcommand{\avg}[1]{\ensuremath{\langle \,#1\, \rangle}}
\newcommand{\etal}{\emph{et al}.}
\newcommand{\delc}{\ensuremath{\delta_c}}
\newcommand{\fnl}{\ensuremath{f_{\rm NL}}}

\newcommand{\eqn}[1]{Eq.~\eqref{#1}}

\newcommand{\fig}[1]{Figure~\ref{#1}}
\newcommand{\figs}[1]{Figures~\ref{#1}}
\newcommand{\ph}[1]{\phantom{#1}}

\newcommand{\be}{\begin{equation}}
\newcommand{\ee}{\end{equation}}

\newcommand{\drm}{\ensuremath{\mathrm{d}}}

\title[Non-Gaussian halo abundances]
      {Non-Gaussian halo abundances in the excursion set approach with correlated
        steps} 

\author[M. Musso \& A. Paranjape]
{Marcello Musso$^{1}$\thanks{E-mail: musso@ictp.it} \& Aseem
  Paranjape$^{1}$\thanks{E-mail: aparanja@ictp.it} \\  
 $^1$ The Abdus Salam International Center for Theoretical Physics,
  Strada Costiera, 11, Trieste 34151, Italy}
\begin{document}
\pagerange{\pageref{firstpage}--\pageref{lastpage}}

\maketitle 

\label{firstpage}

\begin{abstract}
We study the effects of primordial non-Gaussianity on the large scale
structure in the excursion set approach, accounting for correlations
between steps of the random walks in the smoothed initial density
field. These correlations are induced by realistic smoothing filters
(as opposed to a filter that is sharp in $k$-space),
but have been ignored by many analyses to date. We present analytical
arguments -- building on existing results for Gaussian initial
conditions -- which suggest that the effect of the filter at large
smoothing scales is remarkably simple, and is in fact identical to
what happens in the Gaussian case: the non-Gaussian walks behave as if 
they were smooth and deterministic, or ``completely correlated''. As a
result, the first crossing distribution (which determines, e.g., halo
abundances) follows from the \emph{single-scale} statistics of the
non-Gaussian density field -- the so-called ``cloud-in-cloud'' problem
does not exist for completely correlated walks. Also, the answer from 
single-scale statistics is simply one half that for sharp-$k$ walks.

We explicitly test these arguments using Monte Carlo simulations of
non-Gaussian walks, showing that the resulting first crossing
distributions, and in particular the factor 1/2 argument, are
remarkably insensitive to variations in the power spectrum and the
defining non-Gaussian process. We also use our Monte Carlo walks to
test some of the existing prescriptions for the non-Gaussian first
crossing distribution. Since the factor 1/2 holds for both Gaussian
and non-Gaussian initial conditions, it provides a theoretical
motivation (the first, to our knowledge) for the common practice of 
analytically prescribing a ratio of non-Gaussian to Gaussian halo
abundances. 
\end{abstract}


\section{Introduction}
The detection of primordial non-Gaussianity (NG) (Maldacena 2003;
Acquaviva \etal\ 2003) would serve as a powerful discriminator between
models for seeding curvature inhomogeneities in the early universe
(Babich, Creminelli \& Zaldarriaga 2004; LoVerde \etal\ 2008; Sartoris
\etal\ 2010; Sefusatti 2009; see Desjacques \& Seljak 2010 for a
recent review).
Traces of primordial NG can potentially be found not only in the
cosmic microwave backround  (CMB) radiation (Creminelli \etal\ 2006; 
Komatsu \etal\ 2010), but also in
the late time large scale structure of the universe (Slosar
\etal\ 2008). Both the
abundance of collapsed objects (halos) and correlations in their
spatial distribution are sensitive to the
presence of NG in the initial conditions (see, e.g., Dalal
\etal\ 2008; Matarrese \& Verde 2008; Afshordi \& Tolley 2008).
Collapsed objects and the CMB probe very different
spatial scales, and the study of the former therefore promises to
complement the information gained on primordial NG from the
latter. Since detailed studies of collapse and structure formation
involve expensive numerical simulations, it is interesting to explore
the extent to which one can gain analytical insight into the problem. 

The excursion set approach (Epstein 1983; Bond \etal\ 1991; Lacey \&
Cole 1993) has gone a long way in helping 
to build such insight. This approach simplifies the problem of
structure formation by separating its highly non-linear, dynamical
aspects from its statistical ones. The key idea which the approach
hinges on is the realisation that matter does not move very far while
halos form. This allows the problem of the late time distribution
of matter, to be mapped onto its initial conditions. Whether or not a
structure forms on a given length scale is decided by asking whether
or not the initial density field, smoothed on that scale, lies above a
certain threshold (Press \& Schechter 1974). The density field
performs a random walk as the smoothing scale is changed. The
excursion set ansatz relates the statistics of these random walks to
those of large scale structure (e.g., halo abundances, bias, merger
rates, formation times, etc.), while the dynamics enters through the
choice of the threshold, which acts as an absorbing barrier for the
walks. Most importantly for the present discussion,
the mapping of the problem onto the initial conditions brings
primordial NG into the fold of the excursion set approach.

Halo abundances -- the focus of this paper -- in the excursion set
approach are prescribed in terms of the first crossing distribution
(absorption rate) of a barrier, by the random walks performed by the
smoothed density field. If ${\rm d}n/{\rm d}m$ is the differential
number density of halos with mass in $(m,m+{\rm d}m)$, then
\be
 \frac{m}{\bar\rho}\frac{{\rm d}n(m)}{{\rm d}m}{\rm d}m = f(s)\,{\rm
   d}s. 
\label{massfunc}
\ee
where $\bar\rho$ is the mean matter density of the universe, $s$ is
the variance of the density field on scale $R$ with $m=4\pi\bar\rho
R^3/3$, and $f(s)$ is the first crossing distribution.  
In the context of Gaussian initial conditions, it was recognised early
on (Bond \etal\ 1991) that the first crossing distribution $f(s)$
admits an analytical solution when the field is smoothed with a filter
that is sharp in $k$-space, while the problem is much more involved
with a more realistic choice of filter, such as, e.g., a TopHat in real
space. While Bond \etal\ presented numerical results (which we discuss
below), it is worth noting that Peacock \& Heavens (1990) had already
presented an analytical approximation to the first crossing
distribution of filtered walks, which described the numerical results
of Bond \etal\ remarkably accurately.

Based on \eqn{massfunc}, it certainly appears desirable to calculate a
first cross distribution, rather than perform an analysis analogous to
Press \& Schechter (1974) by only calculating unconditional
probabilities for the smoothed density contrast to be above threshold
(although see below). 
In the context of \emph{non}-Gaussian initial conditions, however,
the technical hurdles involved have meant that several analyses have
relied on extending the original Press-Schechter argument
to the non-Gaussian case (Matarrese, Verde \& Jimenez 2000; LoVerde  
\etal\ 2008; LoVerde \& Smith 2011). Only recently have there been
systematic efforts to tackle the problem using the full excursion set
approach of Bond \etal\ (e.g., Lam \& Sheth 2009; Maggiore \& Riotto
2010; D'Amico \etal\ 2011a), very few of which have explicitly
addressed the issue of non-sharp-$k$ filtering (e.g., Maggiore \&
Riotto 2010, who work with the TopHat filter). 

One issue which has hampered the understanding of non-Gaussian halo
abundances, is the fact that even for Gaussian initial conditions,
\emph{both} types of analyses (Press-Schechter and Bond \etal) are
known not to describe well the results of $N$-body simulations. 
In part this is because the exponential tail of the halo 
mass function is not well described by the canonical threshold value 
$\delc=1.686$ from spherical collapse (Gunn \& Gott 1972), a point we
will return to later.
The standard practice in the literature has therefore been to use
Press-Schechter/sharp-$k$ analyses to prescribe a 
\emph{ratio} $R_{\rm NG} = f_{\rm NG}/f_{\rm gauss}$ (e.g. LoVerde
\etal\ 2008). As yet, there has been no convincing explanation of why
such a prescription can be expected to work, although $N$-body
simulations with non-Gaussian initial conditions indicate that the
prescription \emph{does} work (Grossi \etal\ 2007; Pillepich, Porciani
\& Hahn 2008; Desjacques \& Seljak 2010). And finally, it is also
worth keeping in mind that equating the halo abundances on the left
hand side of \eqn{massfunc} with the first crossing distribution on
the right, is only an ansatz (see Paranjape, Lam \& Sheth 2011a,
henceforth PLS11, for a discussion).

Overall then, the present situation as regards the non-Gaussian halo
mass function is, in our opinion, rather clouded. 
There are several
calculations of the function $f_{\rm NG}(s)$ which appears on the
right hand side of \eqn{massfunc}. Accepting that $f_{\rm NG}(s)$ must
be a first crossing distribution, we see that none of the calculations
of $f_{\rm NG}(s)$ have been tested \emph{using} numerical first
crossing distributions. 
Consequently, one cannot claim to understand
which (if any) prescription for $f_{\rm NG}(s)$ works well. 
The ratio prescription has been tested in $N$-body simulations and
seems to do well; however, as yet there is no theoretical
understanding of why this must be so. 
There have been very few attempts to assess the
impact of nontrivial filtering on the calculation of $f_{\rm NG}(s)$,
mainly based on a linearization of the problem (Maggiore \& Riotto
2010). However, since the expansion parameter involved is of order
$\sim0.4$, we would argue that the actual effect of filtering on
non-Gaussian random walks is not yet clear.

Our aim in this paper is to gain some insight into these issues. 
We will therefore focus on testing analytical prescriptions for
non-Gaussian first crossing distributions using Monte Carlo
simulations of random walks (as opposed to mass
functions from $N$-body simulations), along the lines initiated by
PLS11 for the Gaussian case. 
We will demonstrate that the first crossing distribution of sharp-$k$
non-Gaussian walks on the relevant scales is rather insensitive to
the exact nature of the defining non-Gaussian process, and depends,
instead, mainly on the values of the leading order parameters (such as
the skewness) which define the process. 
Our main result, however, concerns the effects of a nontrivial
filter. We will present analytical arguments, supported by our
numerics, to show that for small values of variance $s$ the first
crossing distribution of filtered non-Gaussian walks is simply a
factor $1/2$ times the corresponding distribution for sharp-$k$ walks.
This effect is also remarkably robust against dramatic changes in,
e.g., the power spectrum and type of non-Gaussianity. Since the
same effect is known to be true for the case of \emph{Gaussian} walks,
as an immediate consequence we see that the ratio $f_{\rm NG}/f_{\rm
  gauss}$ of \emph{first crossing distributions} at small $s$ is
independent of the choice of filter. While this does not on its own
explain the effect being seen in simulations, since there is a big
leap involved in going from first crossing distributions to mass
functions, we believe it at least makes the ratio prescription
theoretically plausible. 

The paper is organized as follows. In section 2 we present our
analytical arguments, explaining the role of the filter. 
Section 3 deals with the numerical solution; we
show results for different choices of power spectra and non-Gaussian
processes.
A final section summarizes and discusses some
avenues for further work. The Appendix contains some technical
details regarding the structure of non-Gaussianity.

\section{Analytical arguments: a factor of $1/2$}
The problem of evaluating the first crossing distribution of random
walks with correlated steps for Gaussian initial conditions, was
solved numerically by Bond \etal\ (1991). In the cosmological context,
it is the linearly extrapolated, smoothed density field \del\ which
performs a random walk in steps of its variance
\begin{equation}
  s(R)=\avg{\del_R^2} = \int \drm k \,k^2 \frac{P_\delta(k)}{2\pi^2}
  W^2(kR) \;, 
\end{equation}
where $P_\delta(k)$ is the power spectrum of matter fluctuations,
linearly evolved to present day, and $W(kR)$ is a smoothing filter.
While a filter that is sharp in $k$-space leads to walks with
uncorrelated steps, which can be treated analytically, more realistic
filters such as the TopHat (in real space) or Gaussian lead to
nontrivial correlations between steps. Among other things, Bond
\etal\  demonstrated that the high-mass or small $s$ tail of the first
crossing distribution $f(s)$ is very well described,
\emph{independently} of the choice of filter and power spectrum, by
the Press-Schechter expression
\be
sf_{\rm PS}(s) = \frac12 \frac\delc{\sqrt{2\pi s}} e^{-\delc^2/2s}\,, 
\label{fPS}
\ee
which is a factor $1/2$ different from the analytical answer for walks
with uncorrelated steps. The analytical approximation presented by
Peacock \& Heavens (1990) also reduces to this expression for small
$s$, independently of the filter and power spectrum\footnote{The choice 
of filter does affect how long the small-$s$ regime lasts.
As one can see from Fig.~9 of Bond \emph{et~al.~} (see also Paranjape, 
Lam \& Sheth 2011a), the completely correlated answer is accurate at
least up to $\sqrt{s}\simeq0.5\,\delta_c$ for both TopHat and Gaussian
filters.}.  
In both cases, the actual details of
the filter and power spectrum only come into play when the relation
between $s$ and the smoothing scale $R$ is needed.

This behaviour can be understood in terms of what PLS11 recently
discussed as the 
``completely correlated'' limit of the random walks. In this limit,
the walks are deterministic in the sense that each walk has a constant
value of $\nu=\del/\sqrt{s}$. In the plane of $\del/\sqrt{s}$
vs. $s$, each walk therefore crosses the barrier $\del=\delc$ at most
once. The ``survival probability'', i.e. the fraction of walks that
has not crossed at $s$, is therefore simply the fraction of walks for
which $\nu<\delc/\sqrt{s}$, which is $(1+{\rm
  erf}(\delc/\sqrt{2s}))/2$ for Gaussian initial conditions. The
derivative of this survival probability leads to the Press-Schechter
expression \eqref{fPS}, which is therefore the first crossing
distribution for walks with completely correlated steps and Gaussian
initial conditions.  

In other words, the so-called cloud-in-cloud problem which Bond
\etal\ solved for the sharp-$k$ case, does not exist for completely
correlated walks; such walks either cross the
constant barrier exactly once, or not at all. For correlations
induced by realistic filters (such as the TopHat or Gaussian),
\eqn{fPS} remains a good description at small $s$, since the walks
have not had enough ``time'' to start feeling the effects of
stochasticity; in the small $s$ limit, the walks are still completely
correlated.  

Now consider the case of non-Gaussian initial conditions. In what
follows it will be useful to define the normalised equal-scale
connected moments of the non-Gaussian field \del, $\ep_j \equiv 
\avg{\del^{j+2}}_c/s^{(j+2)/2}$, the most important for us being the
three-point function given by $\ep_1$,
\be
\ep_1 \equiv \frac{\avg{\del^3(s)}}{s^{3/2}}\,.
\label{ep1def}
\ee
A recent analysis by Maggiore \& Riotto (2010) has explored TopHat
filter effects in the calculation of the first crossing
distribution, using a path-integrals analysis. In contrast, we will be
guided by the discussion above of the case of Gaussian initial
conditions. In particular, for small $s$ that entire argument goes
through as is for the non-Gaussian case as well, apart from the fact
that, at any given $s$, the fraction of walks $P_{\rm
  NG}(<\delc/\sqrt{s})$ with $\nu<\delc/\sqrt{s}$ is no longer simply
an error function, but receives non-Gaussian corrections. These
corrections, however, can be calculated simply by knowing the
\emph{single}-point statistics of the non-Gaussian filtered field
$\del$, even though this field in general has a complicated structure
of unequal-scale correlations. This happens because, as far as the
tail of the first crossing distribution is concerned, any memory
induced by the non-Gaussianity is rendered superfluous by these
filter-induced strong correlations. We check this explicitly in the
next section, by numerically calculating the first crossing
distribution of non-Gaussian walks.   

To summarize, we have argued that in its small $s$ tail the first
crossing distribution of walks with non-Gaussian initial conditions
and filter-induced correlated steps must be the derivative of the
survival probability for walks with completely correlated steps:
\be
sf_{\rm NG}(s) = -s\frac{\partial}{\partial s}P_{\rm surv}(s) \to -s 
\frac{\partial}{\partial s} P_{\rm NG}(<\delc/\sqrt{s})\,.
\label{fNG}
\ee
We emphasize that the final result is valid only for small $s$; the
result for Gaussian initial conditions (Peacock \& Heavens 1990; Bond
\etal\ 1991) suggests that the shape of the distribution at large $s$
could be very different. This is not a major concern, since the effect
of primordial non-Gaussianity is expected to be large only at small
$s$, and consequently we need not worry about deviations from this
simple result at large $s$.  

This final result is not new in itself; the single-point
statistics of the non-Gaussian field have been the basis of the
calculations of, e.g., Matarrese, Verde \& Jimenez (2000) (henceforth
MVJ00), LoVerde \etal\ (2008) and LoVerde \& Smith (2011). However,
since those analyses ultimately aimed at obtaining a \emph{mass}
function rather than a first crossing distribution, the true
statistical meaning of those calculations has become clouded, in our
opinion, and is clarified by our arguments above. 

It is worth comparing the structure of the sharp-$k$ non-Gaussian
first crossing distribution with that of the unconditional probability
$P_{\rm NG}(<\delc/\sqrt{s})$ in some detail, since this yields some
insight into the ratio prescription mentioned in the Introduction.
The calculation of Maggiore \& Riotto (2010) shows that the effect of
unequal scale correlations of the non-Gaussian field \del\ in the
first crossing distribution, are typically suppressed by powers of
$\nu^{-1}=\sqrt{s}/\delc$ in the small $s$ tail. As we explain below,
if we therefore ignore these contributions altogether, then 
the derivative with respect to $s$ of the probability
$P_{\rm NG}(<\delc/\sqrt{s})$, \emph{when expressed in terms of the
  equal-scale connected moments $\ep_j$}, will be exactly $1/2$ times 
the sharp-$k$ first crossing distribution,
\be
-s \frac{\partial}{\partial s} P_{\rm NG}(<\delc/\sqrt{s}) = \frac12
\times sf_{{\rm NG,sharp-}k}(s) \,.
\label{factorhalf}
\ee
The reason is that in each
case, sharp-$k$ as well as completely correlated, the non-Gaussianity
appears as an exponentiated derivative operator $\sim
\exp[-\ep_1\p_\nu^3/6+\ldots]$ acting on the corresponding Gaussian
first crossing distribution. This can be seen, e.g., by comparing
equation~(A3) of Paranjape, Gordon \& Hotchkiss (2011) (henceforth
PGH11) for the sharp-$k$ case, with equations~(26) and~(36) of MVJ00
for the completely correlated case.

The theoretical effort now lies in computing the effect of this
derivative operator, and different groups have approached the problem
with different assumptions (see e.g. MVJ00, LoVerde 
\etal\ 2008, Maggiore \& Riotto 2010, D'Amico \etal\ 2011a,
PGH11). The bottom-line for us is the same in each case: since 
the \emph{Gaussian} results for the sharp-$k$ and completely
correlated cases differ by a factor of $1/2$, consequently so do the
non-Gaussian answers. An immediate consequence is that at small $s$,
the ratio of non-Gaussian to Gaussian first crossing distributions
$f_{\rm NG}/f_{\rm gauss}$
will be the same regardless of whether the walks were sharp-$k$
filtered or, more realistically, completely correlated.

As with the case of Gaussian initial conditions, the
relation between $s$ and smoothing scale $R$ explicitly depends on the
choice of filter and power spectrum. Additionally now, the moments
$\ep_j$ will also in general be sensitive to these choices. What is
universal though, is the functional form of the right hand side of
\eqn{factorhalf} when expressed in terms of $s$ and the $\ep_j$. 

For later comparison, we note here that the completely correlated
prediction derived from the  PGH11
prescription\footnote{I.e., $1/2$ times their expression for the
  sharp-$k$ $sf(s)$, the latter being their equation~(A9) divided by
  2.} reads (with $\nu\equiv\delc/\sqrt{s}$) 
\begin{align}
sf_{\rm PGH,cc}(s)=&\frac12\frac{\nu}{\sqrt{2\pi(1+\ep_1\nu)}} 
\nonumber\\ 
&\ph{\frac12}\times
\exp\left[\frac{\ep_1\nu
  -  (1+\ep_1\nu)\ln(1+\ep_1\nu)}{\ep_1^2}\right]\,,
\label{PGHcc}
\end{align}
while the corresponding prediction for MVJ00 is 
\be
sf_{\rm MVJ,cc}(s) = \frac12\frac{\nu}{\sqrt{2\pi(1-\frac13\ep_1\nu)}}
  e^{-\frac12\nu^2(1-\frac13\ep_1\nu)} \left(1-\frac12\ep_1\nu\right)
\,,
\label{MVJcc}
\ee
in which we assumed $\ep_1$ to be constant, which is what
we will use for our numerical results below.

\section{Monte Carlo tests}

To verify that filter-induced correlations do in fact completely
correlate the walks at small $s$, in this subsection we numerically
calculate the first crossing distributions for non-Gaussian random 
walks, comparing the effect of sharp-$k$ and Gaussian filters. The
results are not expected to change when switching to the TopHat
filter; we choose the Gaussian filter since it allows us to derive
simple analytical expressions for some of the quantities we will need
below. Along the way we will also demonstrate that, for the sharp-$k$
walks as well, the first crossing distribution is rather insensitive
to the details of the unequal scale correlations of the non-Gaussian
field.

\subsection{Gaussian initial conditions}

\begin{figure}
 \centering
 \includegraphics[width=\hsize]{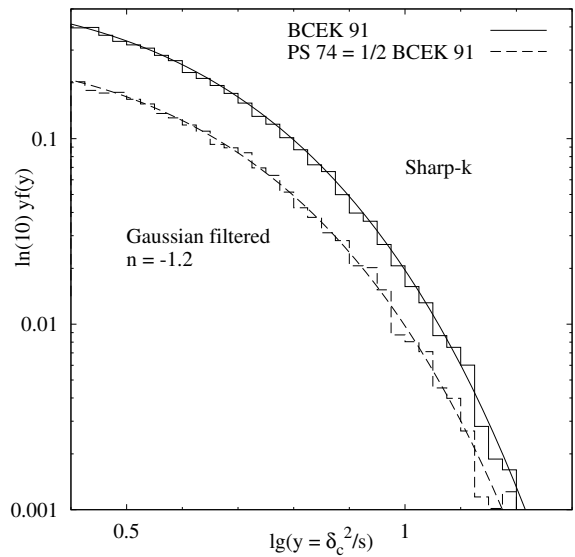}
 \caption{Distribution of scale $s$ at which a walk first crossed the
   constant barrier $\del=\delc$ (histograms), for sharp-$k$ (solid)
   and Gaussian filtered (dashed) walks with Gaussian initial
   conditions. While the filtered walks assumed a power law power
   spectrum $P_\del(k) \propto k^{-1.2}$, the result is insensitive to
   the choice of $P_\del(k)$. The histograms agree with the analytical
   predictions (smooth curves) of Bond \etal\ (1991) (solid) and
   \eqn{fPS} (dashed), respectively.} 
 \label{fig:vfvkGn1p2}
\end{figure}

To set the stage, let us begin by discussing the case of Gaussian
initial conditions in some detail.
The strategy for generating walks with
Gaussian initial conditions is the same as that employed by Bond
\etal\ (1991): the sharp-$k$ walks are constructed by accumulating
independent Gaussian random numbers $g_i$ with a fixed variance
$\Delta s$, $\del_j^{({\rm sharp-}k)} = \sum_{i=1}^jg_i$, while the
Gaussian filtered walks are obtained by applying the filter 
$W(kR)=e^{-(kR)^2/2}$ to the same set of numbers to get
$\del_j^{\rm(Gau)}=\sum_ig_iW(k_iR_j)$. In the latter case, one also
needs to know which values of $k_j$ and $R_j$ to associate to the 
$j$-th step.
Once a power spectrum is specified, this can be done by inverting
the relations $j \Delta S = (2\pi^2)^{-1}\int_0^{k_j} \drm k \,k^2 
P_\delta(k)$ and $j \Delta S = (2\pi^2)^{-1} \int_0^{+\infty} \drm k \,
k^2 P_\delta(k) W^2(k R_j)$.

In this work, we will use a power law form $P_\delta(k)\propto
k^n$ for the power spectrum, rather than the $\Lambda$CDM form, in
order to simplify the analysis. The results of Bond \etal, and our
previous analytical arguments, suggest that the final result for the
non-Gaussian first crossing distribution should be independent of the
form of the power spectrum. We will also explicitly verify this
expectation by displaying results for two very different values of
$n$. As a sanity check, the histograms in \fig{fig:vfvkGn1p2} show our
numerical solution for the first crossing distribution of the constant
barrier $\del=\delc$ for walks with Gaussian initial conditions, for
the sharp-$k$ and Gaussian filters respectively, with $n=-1.2$. The
corresponding smooth curves show the analytical expressions of Bond
\etal\ (1991) and \eqn{fPS}, respectively, which differ by a factor
$1/2$. The agreement indicates that our code is working correctly.

\subsection{Non-Gaussian initial conditions: model definitions}
To generate non-Gaussian walks, we must choose a type of non-Gaussianity. 
Ideally, we would pick as the initial density field a realization of a 
non-Gaussian process compatible with the bispectrum shape of some specific 
model of inflation, and smooth it with the filter of choice to obtain the 
walks.
For example, in the so-called local model (Komatsu \& Spergel 2001;
Lyth, Ungarelli \& Wands 2003; Bartolo, Matarrese \& Riotto 2004;
Dvali, Gruzinov \& Zaldarriaga 2004) the primordial 
curvature perturbation $\zeta(\vec{x})$ is written in real space as $\zeta 
\sim \zeta_G + \fnl\zeta_G^2$, where $\zeta_G$ is a Gaussian random field.
The density contrast \del\ is then obtained from $\zeta$ applying the Poisson
equation and some matter transfer function.

In practice, however, the nonlinear part of this \del\ involves a
convolution in Fourier space over the modes of a Gaussian field
$\delta_G$, which is somewhat cumbersome to handle. 
Since our goal is mainly to demonstrate the effect of the filter (namely, 
the factor of $1/2$ in the first crossing distribution), it is sufficient 
to adopt a simpler model. Our main requirement will be that our model
must reproduce an equal-scale three point function $\ep_1$ of our
choice.
In the local model, $\ep_1$ is weakly scale dependent; for our
numerical walks, we will simply choose a constant value for $\ep_1$. 
Additionally, we will also consider specific choices of unequal-scale
behaviour for the three-point function. We discuss the unequal scale
behaviour of the three-point function of the local model, in Appendix
A1. 

We will consider the class of  integrated non-Gaussian
processes defined by
\begin{align}
  \delta(S) &= \delta_G(S) + \frac{\beta (S)}{3}\!
  \int_0^S \!\drm s \frac{\delta_G(S)\dot\delta_G(s) 
  - \avg{\!\delta_G(S)\dot\delta_G(s)\!}}{\gamma(s)},
\label{NGmod1}
\end{align}
where $\delta_G$ is the Gaussian field smoothed with the filter of
choice, the dot is a derivative with respect to the argument $s$,
and $\beta$ and $\gamma$ are
functions that can be tuned to reproduce the desired behavior of the
three-point function. 
As we show in Appendix A2, setting 
\begin{equation}
  \beta(S) = \epsilon_1 \sqrt{S}
  \bigg[2 \int_0^S \!\drm s
  \frac{\avg{\!\delta_G(S)\dot\delta_G(s)\!}}{\gamma(s)}\bigg]^{-1}
\label{beta}
\end{equation}
allows us to match exactly any value of $\ep_1$, for any choice of
filter. This expression for $\beta(S)$ ignores corrections from a term
involving $\ep_1^3$; we have checked that, for the values of $\ep_1$
we consider in this work, including this correction does not affect
any of our numerical results.
For later use, note that for a sharp-$k$ filter, 
$\avg{\!\delta_G(S)\dot\delta_G(s)\!}=1$, while for the Gaussian
filter and a power law $P_\del(k)\propto k^n$, we have
$\avg{\!\delta_G(S)\dot\delta_G(s)\!} =
2^{(n+3)/2}[1+(s/S)^{2/(n+3)}]^{-(n+5)/2}$.
 
Additionally, setting $\gamma(s)=4\sqrt{s}$ approximately
reproduces the unequal scale behavior of
$\avg{\delta(S_1)\delta(S_2)\delta(S_3)}$ of the filtered local model,
and makes the model above a good proxy to study non-Gaussianity of the
local type (see Appendix A2). We emphasize, however, that the details
of the non-Gaussian process are actually expected to be irrelevant for
the first crossing distribution. One could, e.g., pick the function
$\gamma(s)$ so as to generate any other choice of unequal scale
behaviour. Below, we will also display the results of simply setting
$\gamma(s)=1$.

We will refer to the case $\gamma(s)=4\sqrt{s}$ as model A. 
When generating the random walks for this model, we find it more
convenient to analytically perform an integration by parts and bring
the model definition to the form\footnote{Comparing this form with the
approximate description of the local model in Eq. (11) of Afshordi and
Tolley (2008), we see that our model A is in fact quite close to the
local model. Their term involving $\Phi_{pG}\del_{mG}$ can be compared
with our term $\sim\del(S)\int ds s^{-3/2}\del(s)$: at least with
Gaussian filtering, one has $\del(R)\sim\int d^3k k^2\Phi_{\rm\bf
  k}e^{-k^2R^2/2} \sim (\partial/\partial R^2)\Phi(R)$, so that
$\Phi(s)\sim \int ds g(s)\del(s)$ for some suitable choice of
$g(s)$. This is another way of understanding why, as discussed in the
Appendix, the unequal scale behaviour of the three point function of
model A is close to that of the local model.}
\begin{align}
  \delta(S) &= \delta_G(S) + \frac{\ep_1}{6(1+\alpha)}\!
\bigg[\frac{\del_G(S)^2-S}{\sqrt{S}} \nonumber\\
&\ph{\delta_G(S) + \ep_1(1+\alpha)}
+ \del_G(S)\int_0^S \!\frac{\drm
    s}{2s^{3/2}} \del_G(s) - \alpha\sqrt{S}\bigg] \,,
\label{modAintbparts}
\end{align}
where $\alpha=1$ for the sharp-$k$ filter, and for the Gaussian
filter with $P_\del(k)\propto k^n$, $\alpha=(1.52,1.31)$ for
$n=(-1.2,-2)$, which follows from using the expression for
$\avg{\del_G(S)\dot\del_G(s)}$ mentioned earlier, and performing the
resulting integrals. 

The second model we will consider (henceforth model B) is defined by
setting $\gamma(s)=1$ in \eqn{NGmod1} and choosing $\beta$
appropriately,  
\begin{equation}
  \del(s) = \del_G(s) + \frac{\ep_1}{6}
  \left(\frac{\del_G(s)^2-s}{\sqrt{s}}\right)\,, 
  \label{NGmod2}
\end{equation}
(see Appendix~\ref{app:matching} for details). 
This model significantly differs from model A, and hence from the
local model as well, in the structure of the unequal scale three point
function $\avg{\del(s_1)\del(s_2)\del(s_3)}$. 
As we show in the Appendix, in the limit when $s_1\ll s_2\ll s_3$, model B 
has $\avg{\del(s_1)\del(s_2)\del(s_3)} \propto s_1\sqrt{s_2}$ whereas
the local model has $\avg{\del(s_1)\del(s_2)\del(s_3)}_{\rm loc}
\propto \sqrt{s_1}s_2$. 
 
Model B is also interesting because its single-point statistics can be
calculated analytically (see, e.g., Eqs.~2-5 of MVJ00). We do this in
Appendix B and show that the first 
crossing distribution for walks with completely correlated steps (the
analog of Eqs.~\ref{fPS},~\ref{PGHcc} and~\ref{MVJcc}) is 
\be
sf_{\rm NG,\,modB}(s) = \frac12\frac{\delc}{\sqrt{s}} p_{\rm
  modB}\left(\frac{\delc}{\sqrt{s}},\ep_1\right)\,,
\label{modBfcd}
\ee
where, defining $\lambda(\nu,\ep_1) \equiv (\ep_1/3)(\nu+\ep_1/6)$,
\begin{align}
p_{\rm modB}(\nu,\ep_1) &= \frac{1}{\sqrt{2\pi
    (1+2\lambda(\nu,\ep_1) )}}
\exp\left[-\frac9{\ep_1^2} (1+\lambda(\nu,\ep_1))\right]
\nonumber\\   
&\ph{2\pi2\lambda+1}
\times 2\cosh\left(\frac9{\ep_1^2}
\sqrt{1+2\lambda(\nu,\ep_1)}\right)\,. 
\label{modBpdf}
\end{align}
We will use this below to test our claim that single-point statistics
are sufficient to determine the first crossing distribution of
filtered walks.

\subsection{Non-Gaussian walks: sharp-$k$ case}

Our emphasis here is on the factor $1/2$ which arises due to the walks
being completely correlated at small $s$. As such, it would be enough
to simply display the ratio of the first crossing distributions for
Gaussian filtered and sharp-$k$ walks, checking that it is consistent
with being $1/2$, and we do this below. 
We would like to do more, however, and also compare 
various \emph{theoretical} prescriptions for the first crossing
distribution against our numerical results. Additionally, we wish to
emphasize the point that the sharp-$k$ results for the first
crossing distribution are also insensitive to the
non-Gaussian process. We therefore show the results for sharp-$k$
separately in this subsection, with those for Gaussian filtered walks
in the next subsection. 

\begin{figure}
 \centering
 \includegraphics[width=\hsize]{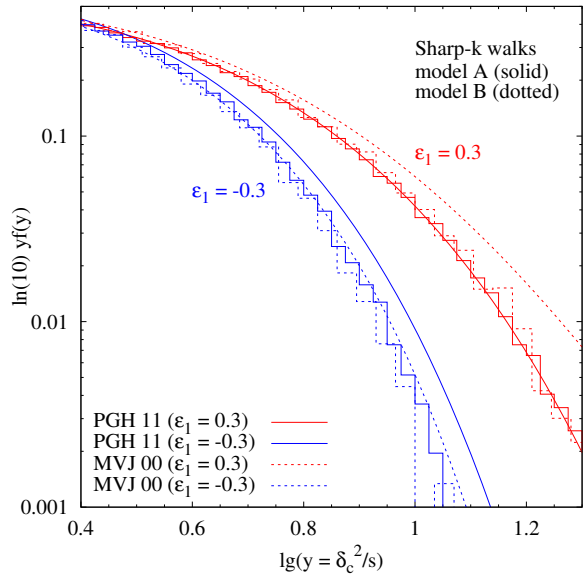}
 \caption{Distribution of scale $s$ at which a walk first crossed the
   constant barrier $\del=\delc$ (histograms), for sharp-$k$ walks
   with non-Gaussian initial conditions defined by models A (solid)
   and B (dotted), for $\ep_1=0.3$ (red, upper) and $\ep_1=-0.3$
   (blue, lower). Note that the histograms for the different models
   are remarkably alike. The smooth curves show the theoretical
   predictions of PGH11 (solid) and MVJ00 (dotted), with the red
   (upper) and blue (lower) curves corresponding to positive and
   negative $\ep_1$, respectively, with $|\ep_1|=0.3$ in each
   case. See text for a discussion.} 
\label{fig:vfvkNGep0p3} 
\end{figure}

\begin{figure}
 \centering
 \includegraphics[width=\hsize]{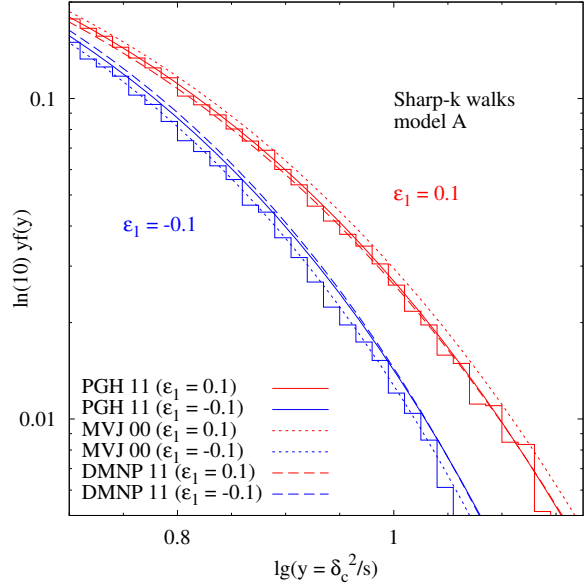}
 \caption{Same as \fig{fig:vfvkNGep0p3}, but for $|\ep_1|=0.1$, with
   histograms shown only for model A. For this case, we also show the
   theoretical predictions of D'Amico \etal\ (2011a) as the dashed
   curves.} 
\label{fig:vfvkNGep0p1}
\end{figure}

\fig{fig:vfvkNGep0p3} shows the results for $|\ep_1|=0.3$, with the
red (upper) and blue (lower) histograms showing the first crossing
distributions of the constant barrier $\del=\delc$, for walks with
positive and negative $\ep_1$, respectively. The solid histograms are
for model A and the dotted for model B. The histograms clearly
demonstrate that, within the numerical accuracy we reach, the
sharp-$k$ first crossing distribution is insensitive to the details of
the non-Gaussian process.

The value $|\ep_1|=0.3$ corresponds to a rather large 
non-Gaussianity, since e.g. in the local model one has $\ep_1\simeq
3\times10^{-4}\fnl$, so that in our case $|\fnl|\sim1000$. 
While we made this choice to visually enhance the effect
of non-Gaussianity in the first crossing distribution, it
also raises an interesting question regarding comparison with
theory. As discussed elsewhere (see e.g. D'Amico \etal\ 2011a and
PGH11), most theoretical analyses rely on a
perturbative expansion in parameter combinations such as $\ep_1\nu$, 
and the large value $|\ep_1|=0.3$ means that these analyses break down 
around $\nu\lesssim3$. The model presented in PGH11,
on the other hand, was specifically constructed to remain well-defined 
in the tail $\nu\gg1$ for $\ep_1>0$, and is therefore a natural choice 
for comparison.

The solid curves show the theory predictions from PGH11 (i.e., $2$
times the expression in \eqn{PGHcc}) for 
positive (red, upper) and negative (blue, lower) $\ep_1$,
respectively.
We see that this prescription performs remarkably
well when $\ep_1>0$, while it does not do well for $\ep_1<0$. The
latter is not surprising, given that the theoretical derivation of
PGH11 strictly works only for $\ep_1>0$, with the $\ep_1<0$
prescription being \emph{ad hoc}. 
The agreement for $\ep_1>0$, on the other hand, highlights once more
the insensitivity of the first crossing distribution to the details of
the non-Gaussianity. This is because the calculation of PGH11 assumed
the  relations $\ep_n=(\ep_1)^n$, whereas in 
our models A and B one can check\footnote{Model
  A has a formal pathology, in that $\ep_2$ logarithmically diverges;
  in terms of a cutoff at some $s=S_0$, the relation at leading order
  in $\ep_1^2$ is $\ep_2=\ep_1^2(26/18+\ln(S/S_0)/6)$. Our numerical
  solution has a natural cutoff $S_0=\Del s$, although in principle
  one can regulate the divergence more carefully. For our walks, the
  logarithmic term at most contributes at the same order as the
  constant $26/18$. We chose not to be more elaborate with the
  regularization, since model B shows
  results identical to model A, without having any such
  pathology; in model B, at leading order we find
  $\ep_2=(4/3)\ep_1^2$. Both models have $\ep_2\neq\ep_1^2$, and
  demonstrate the robustness of the first crossing distribution
  against changes in the exact form of this relation.} e.g. that
$\ep_2\neq\ep_1^2$.  

\begin{figure*}
 \centering
 \includegraphics[width=.49\hsize]{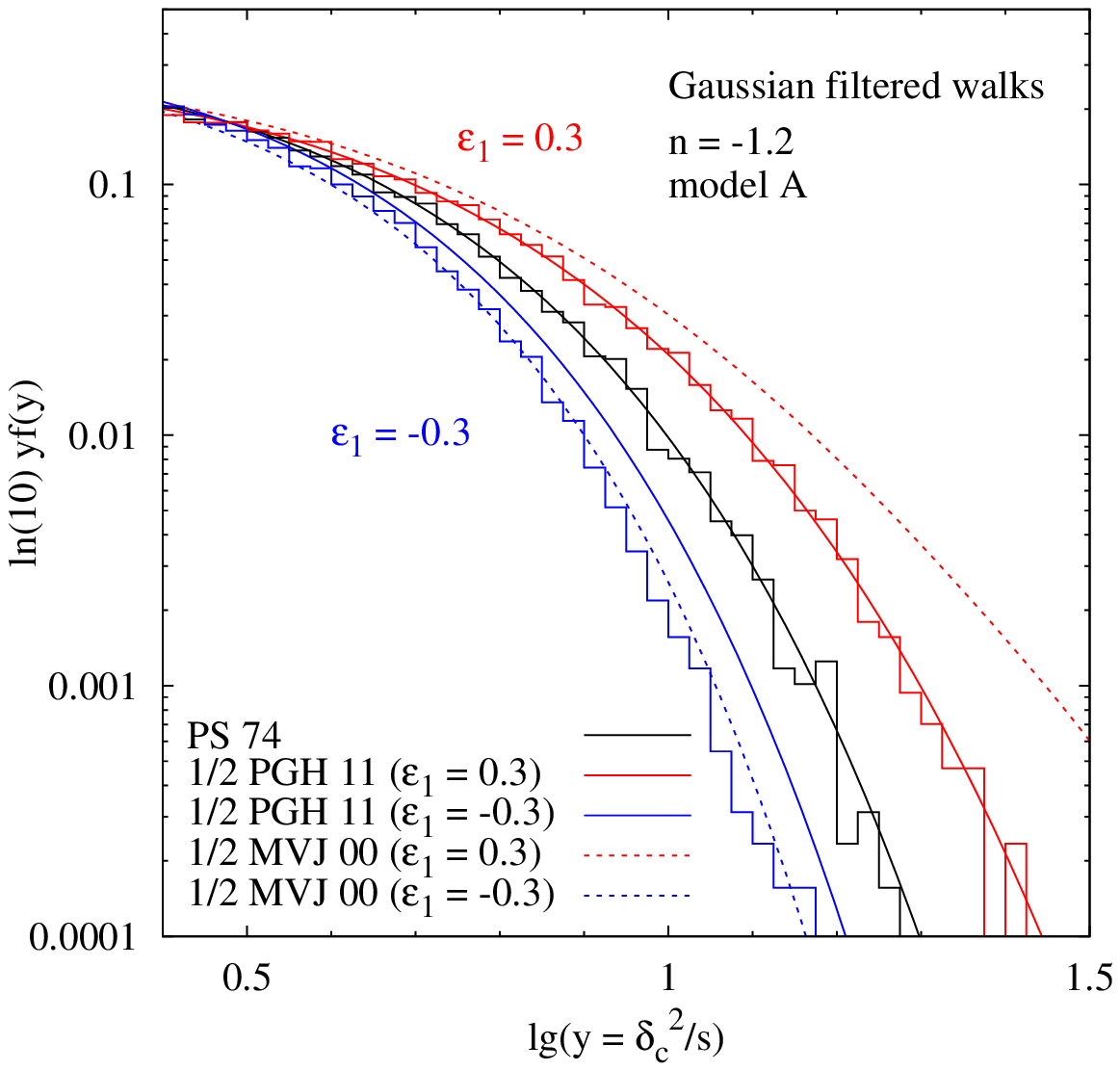}
 \includegraphics[width=.49\hsize]{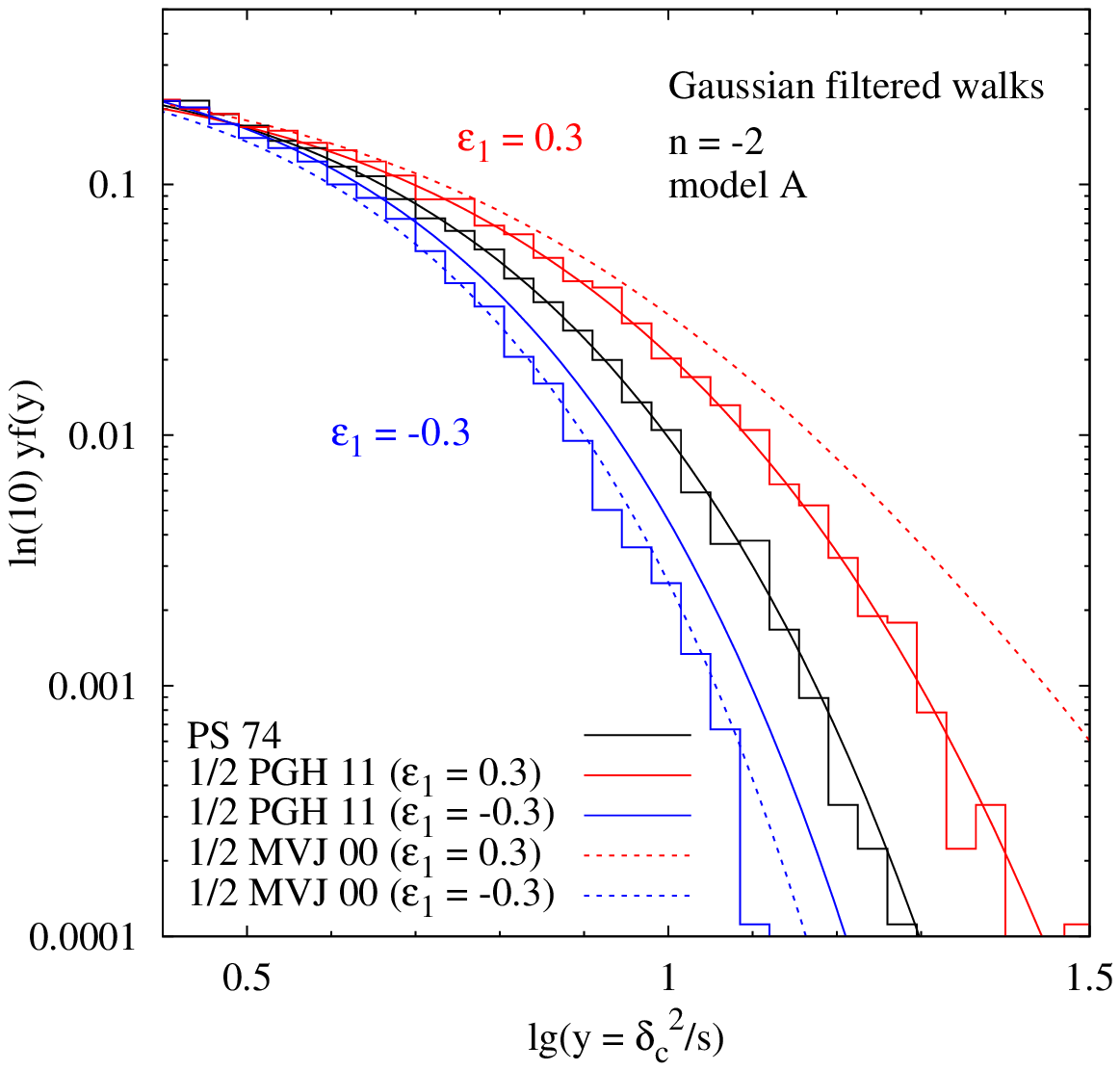}
 \caption{First crossing distribution for Gaussian
   filtered walks having non-Gaussian initial conditions specified by
   model A (histograms) and with $P_\delta\propto k^{n}$, with
   $n=-1.2$ (left panel) and $n=-2$ (right panel). The red (upper) and
   blue (lower) histograms show results for $\ep_1=0.3$ and
   $\ep_1=-0.3$, respectively. For comparison, the black (middle)
   histograms show the first crossing distribution for corresponding
   walks with Gaussian initial conditions. The smooth curves are the
   same in each panel. The black solid (middle) curve is the Gaussian
   prediction \eqn{fPS}. The other solid curves are the non-Gaussian
   prediction \eqn{PGHcc}, with the red (upper) and blue (lower)
   corresponding to $\ep_1=0.3$ and $\ep_1=-0.3$,
   respectively. The dotted curves similarly show the prediction
   \eqn{MVJcc}. See text for a discussion.}
\label{fig:NGmodA} 
\end{figure*}

The dotted curves in \fig{fig:vfvkNGep0p3} show the prediction
of MVJ00 (i.e., $2$ times \eqn{MVJcc}). Given the
large value of $|\ep_1|$, \emph{a priori} one does not expect this
prescription to be a good description of the walks, for either sign of
$\ep_1$. This is quite clearly the case for $\ep_1>0$, with the dotted
curve overestimating the numerical answer by a factor of $4$ or
more. Remarkably, though, the MVJ00 prescription works
quite well for $\ep_1<0$, although it slightly overestimates the
numerical answer.

\fig{fig:vfvkNGep0p1} has the same format as \fig{fig:vfvkNGep0p3},
and shows the sharp-$k$ results for a smaller value $|\ep_1|=0.1$. 
This time, in addition to PGH11 and MVJ00, we
also show the calculation of D'Amico \etal\ (2011a) (dashed
lines). We see that all three prescriptions perform reasonably well
now, although once again the MVJ00 prescription seems to
work better than the others at negative $\ep_1$. 
It is not clear to us whether this is indicating something deeper, or
is simply a coincidence. 

\subsection{Non-Gaussian walks: Gaussian filtered case}

We now turn to the more interesting case of filtered walks. 
\fig{fig:NGmodA} has a similar format as the previous figures, and
shows our results for Gaussian filtered model A
walks, for the first crossing of the constant barrier
$\del=\delc$ . The red (upper) and blue (lower) histograms are for
$\ep_1=\pm0.3$ as before, while the black (middle) histogram shows the
answer for Gaussian initial conditions, with the corresponding
analytical prediction from \eqn{fPS}. The two panels are for
$n=-1.2$ and $-2$, and clearly the corresponding histograms do not
show any significant difference from each other, highlighting the
point we made earlier regarding the irrelevance of the form of the 
power spectrum for the completely correlated tail. We expect the same
to remain true for $\Lambda$CDM as well. 

The theory curves in each panel are the same; the solid curves show
\eqn{PGHcc} (red, upper for $\ep_1=0.3$ and blue, lower for
$\ep_1=-0.3$), while the dotted curves similarly show
\eqn{MVJcc}. Notice that these curves describe the histograms in each
panel as well (or poorly) as the corresponding curves in
\fig{fig:vfvkNGep0p3} described the sharp-$k$ non-Gaussian walks. This is
a clear demonstration that the factor $1/2$ argument works extremely
well in the tail of the distribution, independently of choice of power
spectrum. 

\begin{figure}
 \centering
 \includegraphics[width=\hsize]{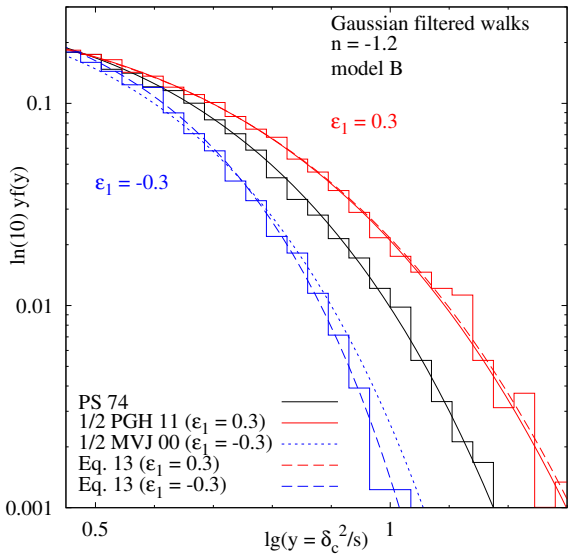}
 \caption{Same as \fig{fig:NGmodA}, but for non-Gaussian walks defined
   by model B, with $n=-1.2$. For the theory curves, we only display
   the $\ep_1>0$ prediction for \eqn{PGHcc} (solid red, upper) and the
   $\ep_1<0$ prediction for \eqn{MVJcc} (dotted blue,
   lower). Additionally, the dashed curves show the prediction of
   single-point statistics for model B in \eqn{modBfcd} for positive
   (red, upper) and negative (blue, lower) $\ep_1$, respectively, with
   $|\ep_1|=0.3$. See text for a discussion.}
\label{fig:NGmodB}
\end{figure}

The histograms in \fig{fig:NGmodB} show the results for Gaussian
filtered model B walks with $n=-1.2$. For the theory curves, we only
display \eqn{PGHcc} for $\ep_1>0$ (red, solid) and \eqn{MVJcc} for
$\ep_1<0$ (blue,dotted). Additionally, the dashed lines show the
prediction in \eqn{modBfcd} of single-point statistics for model
B. The dashed lines compare extremely well with the histograms, and
confirm our argument that single-point statistics are sufficient to
describe the first crossing distribution of filtered walks. The good
agreement between the dashed lines and the other theory curves
indicates that the factor $1/2$ argument is also working well, and is
moreover independent of the type of non-Gaussianity. To make this
point explicitly, \fig{fig:fachalf} shows the ratio of the (numerical)
first crossing distributions for the Gaussian filtered and sharp-$k$
walks. We see that the factor $1/2$ argument works well within our
numerical accuracy. There is an order $\sim10\%$ discrepancy at larger
$s$ for the case of $n=-2$, model A (middle panel),
indicating the limit until which the walks can be assumed to be
completely correlated.

\section{Discussion}
\label{discuss}

An essential ingredient in the application of the excursion set
approach to study structure formation from non-Gaussian initial
conditions, is the ability to calculate the first crossing
distribution $f_{\rm NG}(s)$ of a chosen barrier, for the random walk
performed by the non-Gaussian initial density field $\del(s)$ smoothed
on a scale corresponding to variance $s$. We pointed out that,
while there exist several prescriptions in the literature to calculate
$f_{\rm NG}(s)$ theoretically, as yet there has been no systematic
test of these prescriptions \emph{using} numerically generated first
crossing distributions. We have performed such tests in this work, and
in the process have gained some insight into the structure of $f_{\rm
  NG}(s)$. We summarize our results below:
\begin{itemize}
\item By simulating two models of non-Gaussianity (NG)
  (Eqs.~\ref{modAintbparts} and~\ref{NGmod2}), we argued that for the
  case of sharp-$k$ filtering, $f_{\rm NG}(s)$ is insensitive to the exact
  details of the non-Gaussian process. We also showed that for small
  values of reduced skewness $|\ep_1|$, three different theoretical
  prescriptions (PGH11, MVJ00 and
  D'Amico \etal\ 2011a) work reasonably well (\fig{fig:vfvkNGep0p1}),
  while for larger values of $|\ep_1|$, the PGH11 prescription works
  well for $\ep_1>0$ while the MVJ00 prescription does better for 
  $\ep_1<0$ (\fig{fig:vfvkNGep0p3}). 
\item For the more interesting case of filtered walks, we argued
  that the small $s$ tail of the first crossing distribution is
  essentially determined by the single-point statistics of the
  smoothed non-Gaussian density field. For one of our non-Gaussian
  models (model B) which admits a closed form analytic expression for
  the single-point PDF, we also explicitly verified this numerically
  (\fig{fig:NGmodB}). 
\item We further argued, and demonstrated numerically
  (\figs{fig:NGmodA} and~\ref{fig:NGmodB}), that the first crossing 
  distribution for filtered non-Gaussian walks in the limit of small
  $s$, is simply a factor $1/2$ times the corresponding distribution
  for sharp-$k$ walks. (For the numerical results, we used Gaussian 
  filtering, but the results are not expected to change for TopHat
  filtering.) Since walks with Gaussian initial
  conditions show the same effect, this implies that the ratio $f_{\rm
  NG}/f_{\rm gauss}$ is independent of choice of filter, providing a
  theoretical motivation for the practice of prescribing such a ratio
  when dealing with non-Gaussian mass functions.
\item We demonstrated that the previous result is remarkably robust
  against changing not only the power spectrum (which might have been
  expected based on results for Gaussian initial conditions), but also
  the details of the non-Gaussian process. While this probably means
  that halo abundances on their own will never be very sensitive to
  differences in types of NG, it does also reassure us
  that theoretical calculations can safely make assumptions regarding
  the non-Gaussian process (e.g., the PGH11
  assumption of $\ep_j=(\ep_1)^j$, or simply the single-point
  statistics of model B in \eqn{modBfcd}), without significantly
  affecting the final answer.

\begin{figure}
 \centering
 \includegraphics[width=\hsize]{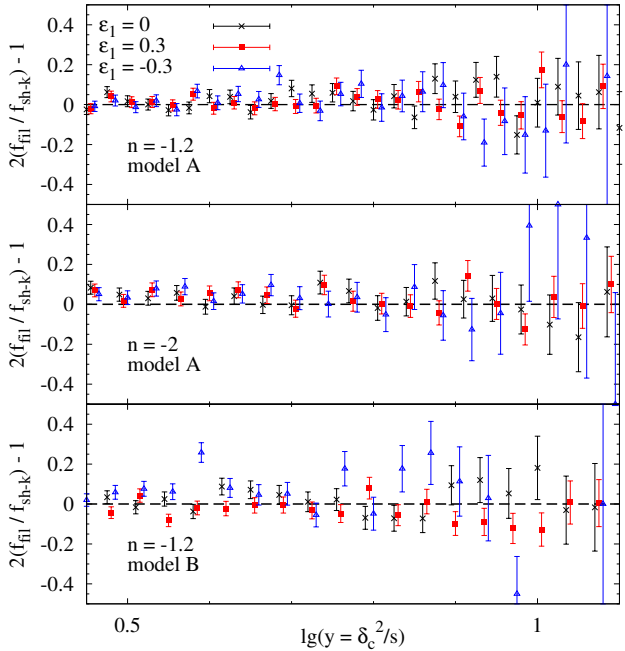}
 \caption{Behaviour of the ratio $f_{\rm
     filtered}/f_{{\rm sharp-}k}$ of Gaussian filtered and sharp-$k$
   first crossing distributions, relative to the predicted value of
   $1/2$. The three panels show various combinations of power spectrum
   $P_{\del}(k)\propto k^n$ and non-Gaussian process as discussed in
   the text. The black crosses, red squares and blue triangles show
   the results (with Poisson errors) for $\ep_1=0$, $0.3$ and $-0.3$, 
   respectively.
} 
\label{fig:fachalf}
\end{figure}

\vskip 0.1in
These considerations imply that studying primordial NG in
the framework of the excursion set approach is actually much simpler
than sharp-$k$ analyses seem to suggest: at small $s$, the first
crossing distributions of filtered walks (which are expected to be
more realistic than sharp-$k$ walks) are determined entirely by the 
single-scale statistics of the initial density field, with no
complications due to the cloud-in-cloud problem. To the extent that
first crossing distributions are sufficient to study halo abundances,
we would argue that, for large masses, it is both simpler and more
accurate to study primordial NG using a Press-Schechter-like approach
rather than the full machinery of excursion sets.

However, as mentioned earlier, equating a first crossing distribution
to a halo mass function is only an ansatz.
All our results are subject to the same caveats as in the case of
Gaussian initial conditions, which were discussed by PLS11. We will
therefore not repeat them here, except to note 
that the issue of stochasticity in the barrier height as a function of
scale (or simply a stochasticity in the constant value of \delc) will
affect the functional form of the ratio $f_{\rm NG}/f_{\rm
  gauss}$. E.g., as PGH11 noted, simply prescribing
a ratio with some chosen value of \delc\ is not sufficient, since the
behaviour of the mass function in the tail will then depend on the
choice of \emph{Gaussian} mass function. In particular, not all
combinations of Gaussian mass function and \delc\ values in the ratio
will lead to well-behaved non-Gaussian mass 
functions. Addressing this issue, however, will require a better
understanding of the jump between first crossing distributions and
mass functions.

One issue which we have not addressed in this work is the question of
halo bias in the presence of non-Gaussianities. Paranjape \& Sheth
(2011) recently discussed halo bias in the excursion set framework
with correlated steps, for Gaussian initial conditions. Their results
built on a simple scaling ansatz proposed by PLS11 for writing the
conditional first crossing distribution, which 
was inspired by bivariate Gaussian statistics and was shown to work
rather well when compared with Monte Carlo simulations. It will be 
interesting to see whether a similarly simple ansatz can be developed
in the non-Gaussian case, especially since, as we have seen, the
effect of the filter at least is exactly the same here as in the
Gaussian case. Previous studies of halo bias with non-Gaussianities
suggest that, in this case, the choice of non-Gaussian process should
be far more relevant than it was for the first crossing
distribution. We leave these issues to future work.

Another discussion we did not enter was the problem of voids in the
excursion set approach (Sheth \& van de Weygaert 2004; Kamionkowski, 
Verde \& Jimenez 2008; Lam, Sheth \& Desjacques 2009; D'Amico
\etal\ 2011b). 
For the case of Gaussian initial conditions, Paranjape, Lam \& Sheth
(2011b) recently argued for a formulation of the problem in terms of
two barriers, one of which is constant and negative and the other
linearly decreasing from positive to negative values. 
The first crossing problem requires counting those walks
which first cross the constant barrier, without having crossed the
linear barrier at smaller $s$. In this case, the difference between
sharp-$k$ and filtered walks is dramatic; whereas the
sharp-$k$ case is extremely sensitive to the exact details of the
falling barrier, the filtered case is very robust, and in fact, the
required first crossing distribution is very well approximated by the
completely correlated answer for the \emph{single} constant
barrier, \eqn{fPS}. This latter result in particular is 
encouraging for the case of non-Gaussian initial conditions; we expect
that the void first crossing distribution in the presence of
non-Gaussianities should also be well described by the single point
statistics of the smoothed non-Gaussian density field using a single
(negative) constant barrier.  However, the sensitivity of the
sharp-$k$ result to the details of the problem implies that the ratio
prescription will probably not work well in this case.

As a final remark, we recall that PLS11 showed, for Gaussian initial
conditions, that the first 
crossing distribution of a constant barrier by completely correlated
walks can be trivially extended to the case when the barrier is a
monotonic function $B(s)$.  Almost all that was needed there, was a
replacement $\delc\to B(s)$ in the survival probability.
In the non-Gaussian case, since we are only interested in the
small $s$, completely correlated tail of the first crossing
distribution, and since the arguments of PLS11 did not rely
on the initial conditions being Gaussian, it should be a
straightforward matter to extend the non-Gaussian constant-barrier
results to the case of moving barriers. In particular, this seems to
be a promising way of approaching the problem of re-ionization in the
presence of non-Gaussianity, following, e.g., Furlanetto, Zaldarriaga
\& Hernquist (2004). 

\section*{Acknowledgements}
We thank Ravi Sheth for several insightful discussions, and Sabino
Matarrese and Ravi Sheth for comments on an earlier draft.

\end{itemize}

\appendix

\section{The unequal scale behavior of the three-point function} 
\label{app:matching}
In this Appendix we investigate the behavior of the three-point
function $\avg{\del_1\del_2\del_3}$ where $\del_j=\del(s_j) =
\del(s(R_j))$, in the regime where $R_1\gg R_2\gg R_3$, i.e.~when
$S(R_1) \ll S(R_2) \ll S(R_3)$. 
We begin by discussing models inspired from inflationary physics, in
particular the local model, and then compare with the simpler models
we have considered in this work.

\subsection{The local model}
The three-point function of $\delta(R)$ can always be written in terms
of the bispectrum $B(k_1,k_2,k_3)$ of the non-Gaussian curvature
pertubation $\zeta$ generated during inflation, as 
\begin{align}
  \avg{\delta_1\delta_2\delta_3} &\propto \int\drm k_1 k_1^3 T(k_1)
  W(k_1 R_1) \int\drm k_2 k_2^3 T(k_2) W(k_2 R_2) \notag \\
  &\times \int_{|k_2-k_1|}^{k_1+k_2}\drm y \,y^3 T(y) W(y R_3)
  B(k_1,k_2,y), 
\label{app:3plocal}
\end{align}
where $T(k)$ is the matter transfer function. For a CDM
cosmology, $T(k)$ decays like $T(k)\sim (1/k)^2\log k$ for $k\gg
k_{\rm eq}$, where $k_{\rm eq}$ is the scale corresponding to
matter-radiation equality (Bardeen \etal\ 1986).

With the assumed hierarchy of scales, this expression for
$\avg{\delta_1\delta_2\delta_3}$ is never very  sensitive to
$R_3$. Indeed, because of the filters $W(k_1 R_1)$ and $W(k_2 R_2)$,
the integrand is suppressed much before $k_1+k_2$ becomes of order
$1/R_3$, and one can thus assume $y \ll1/R_3$, for which
$W(yR_3)\simeq 1$. Moreover, since $1/R_2 \gg 1/R_1$, the $k_2$
integral has a much larger range than the $k_1$ one. For $k_2\gg k_1$
the width of the $y$ integral becomes small, and its value can be
approximated by the value of the integrand at $k_2$ times the width,
i.e.~$2k_1 k_2^3 T(k_2) B(k_1,k_2,k_2) W(k_2R_3)$. 

For non-Gaussianity of the local type, the bispectrum gives
$B(k_1,k_2,k_2)\propto 2P_\zeta(k_1)P_\zeta(k_2) + P_\zeta^2(k_2)$,
where $P_\zeta(k)$ is the primordial power spectrum of curvature
perturbations. In the $k_2\gg k_1$ limit, the integrand above thus
reduces to the sum of the two factorized terms 
\begin{align}
  4 &\left[k_1^4 P(k_1) T(k_1) W(k_1 R_1)\right]\! \notag\\
  &\times\left[k_2^6 P(k_2) T^2(k_2) W(k_2 R_2)W(k_2 R_3)\right]\,, 
\label{app:dominant}
\end{align}
and
\begin{align}
  2 &\left[k_1^4 T(k_1) W(k_1 R_1)\right]\! \notag\\
  &\times\left[k_2^6 P^2(k_2) T^2(k_2) W(k_2 R_2)W(k_2 R_3)\right].
\end{align}
For large $k_2$ (and assuming a scale invariant power spectrum), the
first term scales like $(1/k_2)\log^2(k_2)$, which would diverge if it
were not for the filter. The integral over $k_2$ is thus dominated by
the value around $1/R_2$, and is largely insensitive to the exact
behavior of the integrand for $k_2\lesssim k_1$. The second term
instead scales like $(1/k_2^4)\log^2(k_2)$, which is convergent and
subleading. The integral in \eqref{app:3plocal} is therefore well
approximated by the factorized integral of \eqref{app:dominant} over
$k_1$ and $k_2$, the second of which is simply the two-point function  
\begin{equation}
  \avg{\delta_2\delta_3} \propto
  \int \drm k_2 k_2^6 P(k_2) T^2(k_2) W(k_2 R_2)W(k_2 R_3)\;.
\end{equation} 
Also, the integral over $k_1$ is dominated by values around $1/R_1$,
where its integrand (at scales smaller than the equality scale)
behaves like $(1/k_1)\log(k_1)$. Confronting it with the variance
$S(R_1)$, which in the same regime scales like the integral of
$(1/k_1)\log^2(k_1)$, we conclude that the scaling of the three-point
function is 
\begin{equation}
  \avg{\delta(S_1)\delta(S_2)\delta(S_3)} \sim \sqrt{S_1}\;
  \avg{\delta(S_2)\delta(S_3)}\;,
\end{equation}
which reduces to $\sqrt{S_1}S_2$ with a sharp-$k$ filter.

\subsection{Models A and B}
Turning now to the non-Gaussian models used in this paper, the
three-point function of model A defined in \eqn{NGmod1} reads at
linear order in $\beta$  
\begin{equation}
  \avg{\delta_1\delta_2\delta_3} =
  \frac{\beta(S_1)}{3} \int_0^{S_1} \!\!\drm s
  \frac{\avg{\delta_2\delta_1}\avg{\delta_3\dot\delta(s)}}{\gamma(s)}
  + \mathrm{5~perms.};
\end{equation}
from this expression, setting $S_1=S_2=S_3=S$ one gets the skewness
\begin{equation}
  \avg{\delta^3(S)}=2\beta(S) S \int_0^S \!\!\drm s
  \frac{\avg{\delta(S)\dot\delta(s)}}{\gamma(s)}\;,
\end{equation}
and setting $\avg{\delta^3}=\epsilon_1S^{3/2}$, where $\epsilon_1$
is the reduced third moment of the model we want to reproduce, one
immediately recovers \eqn{beta}. 

The equal scale behavior of the three-point function can be matched
exactly by fixing $\beta$, for an arbitrary function $\gamma(s)$. We
now want to use the residual freedom to reproduce (at least
approximately) the unequal scale behavior. 
In the sharp-$k$ case one has $\avg{\delta_i\delta_j}=\min(S_i,S_j)$
and $\avg{\delta_i\dot\delta(s)}=\vartheta(S_i-s)$; setting
$\gamma(s)=4\sqrt{s}$ simply yields 
\begin{equation}
  \avg{\delta_1\delta_2\delta_3} = \frac{\beta_1}{3}S_1^{3/2}
  +\frac{\beta_2+\beta_3}{6}
  \big(S_1\sqrt{S_2}+S_2\sqrt{S_1}\big)
\end{equation}
with $\beta_i=\epsilon_1(S_i)$. We notice that, as long as $\ep_1$ is 
nearly constant, this result is nearly
independent of $S_3$. Secondly, in the $S_1\ll
S_2$ limit the dominant term is the one with $\sqrt{S_1}$ and the
three-point function becomes 
\begin{equation}
  \avg{\delta_1\delta_2\delta_3}\simeq \frac{\epsilon_1}{3}\sqrt{S_1}S_2\;, 
\end{equation}
which matches the approximate scaling of the local model. 

For a generic filter the computations are slightly less immediate, but
one can check that the very same arguments that lead to the previous
formula go through, with $S_2$ replaced by $\avg{\delta_2\delta_3}$,
just like what happens for the local model. 
This matching could in principle be made more accurate by choosing
$\gamma$ so that the exact scaling with $S_1$ of the $k_1$ integral in
\eqn{app:dominant} is reproduced.
However, since our simulations show that the fine
details of the unequal scale behavior of
$\avg{\delta_1\delta_2\delta_3}$ are not very important, for the level
of accuracy requested in this paper we are satisfied with the simple
expression for $\gamma(s)$. 

Finally, the three-point function of the non-Gaussian model B defined
in \eqn{NGmod2}, for a generic filter, reads
\begin{align}
\avg{\del_1\del_2\del_3} &= \frac23 (S_1S_2S_3)^{1/2} \nonumber\\
&\ph{23}\times
\bigg[ \beta_1r_{12}r_{13} +  \beta_2r_{21}r_{23} + \beta_3
  r_{31}r_{32} \nonumber\\
&\ph{23\times\beta_1r_{12}}
+ \frac49(\beta_1\beta_2\beta_3) r_{12}r_{23}r_{31}\bigg]\,,
\label{modB123gen}
\end{align}
where we defined $r_{ij} \equiv \avg{\del_i\del_j}/\sqrt{S_iS_j}$, and
$\beta_i=\beta(s_i)$. For the sharp-$k$ filter, at leading order in
$\beta$, this becomes
\begin{equation}
  \avg{\delta_1\delta_2\delta_3} = \frac{2}{3}
  \bigg[\beta_1 S_1^{3/2} +\beta_2 S_1\sqrt{S_2}
  +\beta_3 \frac{S_1S_2}{\sqrt{S_3}}\bigg]\,.
\end{equation}
Again, setting $S_1=S_2=S_3=S$ and
matching against $\avg{\delta^3}=\epsilon_1S^{3/2}$ yields
$\beta(S)=\epsilon_1/2$. In the limit
$S_1\ll S_2\ll S_3$ the dominant term is the second one: we have
therefore 
\begin{equation}
  \avg{\delta_1\delta_2\delta_3} \simeq \frac{\epsilon_1}{3} S_1\sqrt{S_2}
\end{equation}
which does not match the scaling of the local model in the
hierarchical regime. Similar results follow for generic filters by using
\eqn{modB123gen}.

\section{PDF for Model B}
\label{app:PDFmodB}

The non-Gaussian PDF for $\nu=\delta/\sqrt{S}$ in Model B can be
computed from the Gaussian PDF for $\delta_G$ as 
\begin{equation}
  p(\nu) = 
  \int \drm x \, \frac{e^{-x^2/2}}{\sqrt{2\pi }} 
  \delta_\mathrm{D}\bigg(\nu - x
  - \epsilon\frac{x^2-1}{6}\bigg)\,,
\end{equation}
where $\ep\equiv\ep_1$. One now applies the relation
$\delta_\mathrm{D}[f(x)]=\sum_i\delta_\mathrm{D}(x-x_i)/|f'(x_i)|$
(the $x_i$'s being the roots of the equation $f(x)=0$) for the change
of variables in the Dirac delta function. Here $f(x)=0$ is just a
quadratic equation with the two roots 
\begin{equation}
  x_\pm = \frac{3}{\epsilon} \bigg[
  -1\pm\sqrt{1+\frac{2\epsilon}{3}\bigg(\nu+\frac{\epsilon}{6}\bigg)}\bigg], 
\label{app:roots}
\end{equation}
while its Jacobian gives 
\begin{equation}
  |f'(x_+)| = |f'(x_-)| =
  \sqrt{1+\frac{2\epsilon}{3}\bigg(\nu+\frac{\epsilon}{6}\bigg)} \; ;
\end{equation}
the PDF therefore reads
\begin{equation}
  p(\nu) = \frac{e^{-x_+^2/2}+e^{-x_-^2/2}}{
  \sqrt{2\pi}\sqrt{1+(2\epsilon/3)(\nu+\epsilon/6)}} \;.
\end{equation}
For small $\epsilon$ one has $x_+\simeq\nu$ and
$x_-\simeq-6/\epsilon$; therefore in this limit $e^{-x_-^2/2}$ is
exponentially suppressed, and from $e^{-x_+^2/2}$ one correctly
recovers the Gaussian result. For generic values of $\epsilon$,
plugging in the above expression the roots $x_+$ and $x_-$ from
\eqn{app:roots} gives the result quoted in \eqn{modBpdf}. 
The first crossing distribution for walks with completely correlated
steps follows from constructing the survival probability
$\int^{\delc/\sqrt{s}}{\rm d}\nu p(\nu)$ and then differentiating with
respect to $s$, which leads to the expression in \eqn{modBfcd}.

\label{lastpage}

\end{document}